\documentclass[twocolumn,aps,prl,epsf,showpacs]{revtex4}
\usepackage{amsmath}
\usepackage{epsfig}
\usepackage{epsf}
\usepackage{array}

\begin{document}

\title{Spatial inhomogeneity and strong correlation physics: a dynamical
mean field study of a model Mott-insulator/band-insulator heterostructure}
\author{Satoshi Okamoto and Andrew J. Millis}
\affiliation{Department of Physics, Columbia University, 538 West 120th Street, New York,
New York 10027, USA}
\date{\today }

\begin{abstract}
We use the dynamical mean field method to investigate electronic properties
of heterostructures in which finite number of Mott-insulator layers are
embedded in a spatially infinite band-insulator. The evolution of the
correlation effects with the number of Mott insulating layers and with
position in the heterostructure is determined, and the optical conductivity
is computed. It is shown that the heterostructures are generally metallic,
with moderately renormalized bands of quasiparticles appearing at the
interface between the correlated and uncorrelated regions.
\end{abstract}

\pacs{73.21.-b,71.27.+a,73.40.-c,78.20.-e}
\maketitle



An exciting new direction in materials science is the fabrication and study
of heterostructure involving \textquotedblleft correlated electron
\textquotedblright\ materials such as Mott insulators, high-temperature
superconductors, and novel magnets\cite{Imada98,Tokura00}. The issues raised
by these heterostructures, especially the evolution with position of
properties from correlated to uncorrelated, is of fundamental physical
interest and would be crucial for prospective devices based on correlated
electron compounds. Many interesting systems have been fabricated, including
modulation-doped high-$T_{c}$ superconductors\cite{Ahn99,Ahn02},
Mott-insulator/band-insulator heterostructures\cite{Ohtomo02}, and a variety
of combinations of magnetic transition-metal oxides\cite%
{Izumi01,Biswas00,Biswas01}, but there has been relatively little
theoretical study of the heterostructure-induced changes in many-body
physics. The theoretical problem is difficult because it requires methods
which can deal both with spatial inhomogeneity and strong correlation
physics.

Despite the difficulties, several interesting works have appeared. Fang,
Soloviev and Terakura\cite{Fang00} used bulk band structure calculation to
gain insight into the effects \cite{Izumi01} of strain fields induced by
lattice mismatch in a heterostructure. Matzdorf and co-workers used GGA band
theory methods to study the surface electronic and lattice structure of Sr$%
_{2}$RuO$_{4}$, predicting a lattice distortion which is observed and
surface ferromagnetism which is apparently not observed \cite{Matzdorf00}.
Potthoff and Nolting used dynamical-mean-field methods to study the
consequences of the lower coordination at a surface \cite%
{Potthoff99,Schwieger03}.

All of these papers, however, treated situations in which the electronic
density remained at the bulk value, and the new physics arose from
structural differences. A crucial feature of heterostructures is an
inhomogeneous electron density caused by a spreading of charge across the
interfaces which define the system. Recently \cite{Okamoto04a,Okamoto04b},
we used realistic multi-orbital interaction parameters and a
density-functional-theory-derived tight-binding band structure to model
ground state properties of the LaTiO$_{3}$/SrTiO$_{3} $ heterostructure
fabricated by Ohtomo \textit{et al.}\cite{Ohtomo02}. \ While this study
captured important aspects of the density inhomogeneity, it did not address
the dynamical properties of correlated heterostructures. Further, this study
employed the Hartree-Fock approximation which is known to be an inadequate
representation of strongly correlated materials, and in particular does not
include the physics associated with proximity to the Mott insulating state.

In this paper we use the dynamical mean field method \cite{Georges96}, which
provides a much better representation of the electronic dynamics associated
with strong correlations, to study the correlated electron properties of a
simple Hubbard-model heterostructure inspired by--but not a fully realistic
representation of--the systems studied in \cite{Ohtomo02}. We present
results for observables including photoemission spectra, optical
conductivity, charge density, and highlight similarities and differences to
previous work.

We study a single orbital model with basic Hamiltonian $%
H_{hub}=H_{band}+H_{int}+H_{coul}$ with 
\begin{eqnarray}
H_{band} &=&-t\sum_{\langle ij\rangle, \sigma}(d_{i\sigma }^{\dag }d_{j\sigma
}+H.c.),  \label{Hband} \\
H_{int} &=& U\sum_{i}n_{i\uparrow }n_{i\downarrow } +\frac{1}{2}\sum_{
{i\neq j} \atop {\sigma, \sigma^{\prime}}} \frac{e^{2}n_{i \sigma}n_{j
\sigma^{\prime}}}{\varepsilon |\vec{R}_{i}-\vec{R}_{j}|}.  \label{Hint}
\end{eqnarray}%
Here the sites $i$ form a simple cubic lattice of lattice constant $a$, so
electronic positions $\vec{R}_{i}=a(n_{i},m_{i},l_{i})$. We include both an
on-site ($U$) and long ranged Coulomb interaction: the screening field from
the latter is important for the electron density profile. We associate the
electronic sites with the B-sites of an ABO$_3$ perovskite lattice, and
define the heterostructure by counterions of charge $+1$ placed on a subset
of the A-sites. Here we study an $n$-layer $[001]$ heterostructure defined
by $n$ planes of $+1$ counterions placed at positions $\vec{R}%
_{j}^{A}=a(n_{j}+1/2,m_{j}+1/2,l_{j}+1/2)$, with $-\infty
<n_{j},m_{j}<\infty $ and the $l_{j}$ runing over $n$ adjacent values. The
resulting potential is 
\begin{equation}
H_{coul}=-\sum_{i,j, \sigma}\frac{e^{2}n_{i \sigma}} {\varepsilon |\vec{R}%
_{i}-\vec{R}_{j}^A|}.  \label{Hcoul}
\end{equation}%
Charge neutrality requires that the areal density of electrons is $n$. 
A dimensionless measure of the strength of the Coulomb 
interaction is $E_c=e^2/(\varepsilon a t)$; we choose parameters
somewhat arbitrarily so that $E_c=0.8$ (this corresponds to 
$t \sim 0.3$~eV and length $a \sim 4$~\AA \ and $\varepsilon=15$,
which describe the system studied in \cite{Ohtomo02}). We found
that the charge profile did not depend in an important way on $\varepsilon $
for $5<\varepsilon <25$.

The basic object of our study is the electron Green function, which for the
[001] heterostructure may be written 
\begin{equation}
G(z,z^{\prime },\vec{k}_{\parallel };\omega )=[\omega +\mu
-H_{band}-H_{Coul}-\Sigma (z,z^{\prime },\vec{k}_{\parallel };\omega )]^{-1}.
\label{eq:Greenlatt}
\end{equation}%
We approximate the self-energy operator as the sum of a Hartree term arising
from the long-ranged part of the Coulomb interaction 
\begin{equation}
\Sigma _{H}(z_{i})=\sum_{j\neq i,\sigma }\frac{e^{2} \langle n_{j\sigma}
\rangle}{\varepsilon |\vec{R}_{i}-\vec{R}_{j}|}  \label{sigH}
\end{equation}%
and a dynamical part $\Sigma _{D}$ arising from local fluctuations .
Following the usual assumptions of dynamical-mean-field theory \cite%
{Georges96} as generalized to inhomogeneous situations by Schwieger \textit{%
et. al.} \cite{Schwieger03}, we assume 
\begin{equation}
\Sigma _{D}\Rightarrow \Sigma _{D}(z,\omega ).  \label{sigd}
\end{equation}%
The layer $(z)$-dependent dynamical self energy $\Sigma _{D}$ is determined
from the solution of a quantum impurity model \cite{Georges96} with
mean-field function fixed by the self-consistency condition 
\begin{equation}
G^{imp}(z,\omega )=\int \frac{d^{2}k_{\parallel }}{(2\pi )^{2}}G(z,z,\vec{k}%
_{\parallel };\omega ).  \label{sce}
\end{equation}%
One must solve a separate impurity model for each layer, but the self
consistency condition [cf. Eq.~(\ref{sce})] implies the solutions are
coupled. It is also necessary to self-consistently calculate the charge
density via $n(z)=-2 \int \frac{d\omega }{\pi }f_{\omega }\mathrm{Im}
G^{imp}(z,\omega )$ with $f$ the Fermi distribution function. The numerics
are time consuming, and it is therefore necessary to adopt a computationally
inexpensive method for solving the quantum impurity models. We use the
two-site method of Potthoff \cite{Potthoff01}, which reproduces remarkably
accurately the scaling of the quasiparticle weight and lower Hubbard band
near the Mott transition. We have also verified \cite{Okamoto04c} that the
two site method reproduces within a few percent the $T=0$ magnetic phase
diagram found by Ulmke \cite{Ulmke98} in a model with an unusual low-energy
density of states peak.

\begin{figure}[tbp]
\begin{center}
\includegraphics[scale=0.5,clip]{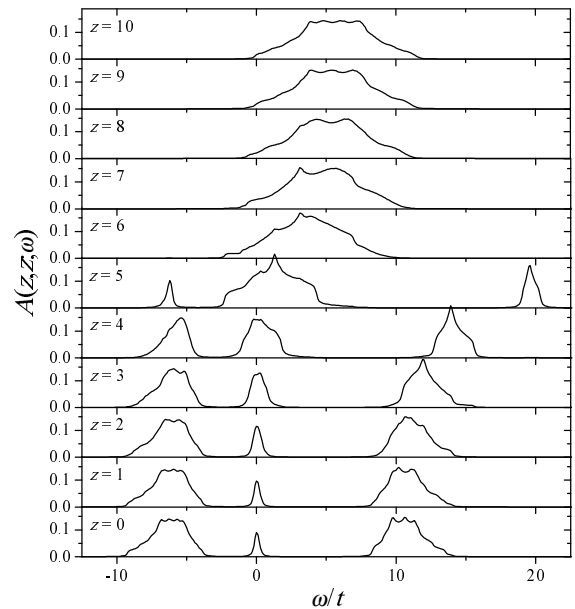}
\end{center}
\par
\caption{Layer-resolved spectral function calculated for $10$-layer
heterostructure for $U=16t$, $\protect\varepsilon =15$. The heterostructure
is defined by +1 charges placed at $z=\pm 0.5,\pm 1.5,...\pm 4.5$ so the
electronic (B) sites are at integer values of $z$. }
\label{fig:spectra}
\end{figure}

Fig.~\ref{fig:spectra} shows the layer-resolved spectral function $%
A(z,z;\omega)=-\frac{1}{\pi }\int \frac{d^{2}k_{\parallel }}{(2\pi )^{2}}%
\mathrm{Im}G(z,z,\vec{k}_{\parallel };\omega +i0^{+})$ for a $10$-layer
heterostructure with $U=16t$ (about 25\% greater than the critical value
which drives a Mott transition in a bulk system described by $H_{hub}$ with $%
n=1$. The spectral functions are in principle measureable in photoemission
or scanning tunneling microscopy. Outside the heterostructure ($z>6$), the
spectral function is essentially identical in form to that of the free
tight-binding model $H_{band}$. The electron density is negligible, as can
be seen from the fact that almost all of the spectral function lies above
the chemical potential. As one approaches the heterostructure ($z=6$), the
spectral function begins to broaden. Inside it ($z\preceq 5$) weight around $%
\omega =0$ begins to decrease and the characteristic strong correlations
structure of lower and upper Hubbard bands with a central quasiparticle peak
begins to form. The sharp separation between these features is an artifact
of the 2-site DMFT [as is, we suspect, the shift in energy of the upper
(empty state) Hubbard band for $z=4,5$]. Experience with bulk calculations
suggests that the existence of three features and the weight in the
quasiparticle region are reliable. Towards the center of the
heterostructure, the weight in the quasiparticle band becomes very small,
indicating nearly insulating behavior. For very thick heterostructures, we
find the weight approaches $0$ exponentially.

\begin{figure}[tbp]
\epsfxsize=0.75\columnwidth \centerline{\epsffile{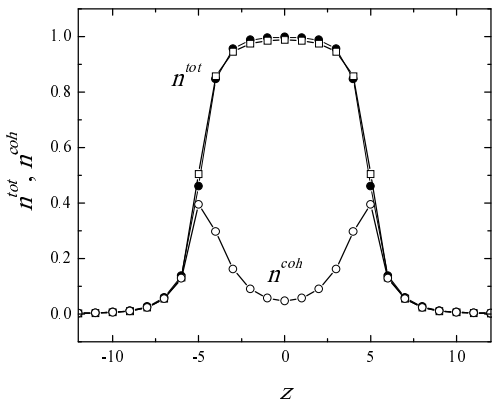}}
\caption{Total charge density (open squares) and charge density from
coherent part near Fermi level (open circles). For comparison, total charge
density calculated by applying Hartree Fock approximation to the Hamiltonian
shown as filled symbols. Parameters as in Fig.~\protect\ref{fig:spectra}.}
\label{fig:density}
\end{figure}

The behavior shown in Fig.~\ref{fig:spectra} is driven by the variation in
density caused by leakage of electrons out of the heterostructure region.
Fig.~\ref{fig:density} shows the numerical results for the charge-density
distribution $n(z)$ for the heterostructure whose photoemission spectra are
shown in Fig.~\ref{fig:spectra}. One sees that in the center of the
heterostructure ($z=0$) the charge density is approximately $1$ per site,
and that there exists an edge region, of about three-unit-cell width, over
which the density drops from $\sim 1$ to $\sim 0$. The over-all charge
profile is determined mainly by the self consistent screening of the Coulomb
fields which define the heterostructure, and is only very weakly affected by
the details of the strong on-site correlations (although the fact that the
correlations constrain $n<1$ is obviously important). To show this, we have
used the Hartree-Fock approximation to recalculate the charge profile: the
results are shown as filled circles in Fig.~\ref{fig:density} and are seen
to be almost identical to the DMFT results.

The existence of an approximately three unit cell wide edge region where the
density deviates significantly from the values $n=0$ and $n=1$
characteristic of the two systems in bulk form implies that only relatively
thick heterostructures ($n>6$) will display `insulating' behavior in their
central layers, and suggests that the edge regions sustain quasiparticle
subbands which give rise to metallic behavior. The open circles in Fig.~\ref%
{fig:density} show the charge density in the `quasiparticle bands' [obtained
by integrating $A(z,z;\omega )$ from $\omega =0$ down to the first point at
which $A(z,z;\omega )=0$]. \ One sees that these near Fermi-surface states
contain a small but non-negligible fraction of the total density, suggesting
that edges should display relatively robust metallic behavior. The results
represent, a significant correction to the Hartree-Fock calculation \cite%
{Okamoto04b}, which leads, in the edge region, to a metallic quasiparticle
density essentially equal to the total density.

\begin{figure}[tbp]
\begin{center}
\includegraphics[scale=0.45,clip]{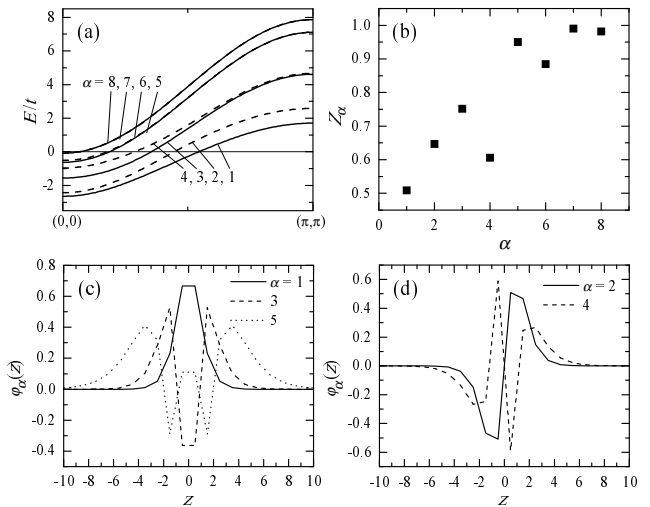}
\end{center}
\par
\caption{(a) Dispersion relations of filled-subband quasi particles
calculated for $3$ layer heterostructure with $U=16t$ and $\protect%
\varepsilon=15$. Solid and broken lines are for odd- and even-$\protect%
\alpha $, respectively. (b) Subband quasiparticle weights. (c, d)
Quasiparticle wave functions for $\protect\alpha=1,3,5$ and 2,4 at $\protect%
\omega=0$. Here the heterostructure is defined by +1 charges placed at $z=0,
\pm 1$ so the electronic (B) sites are at half integer values of $z$.}
\label{fig:QP}
\end{figure}

The spectral function is determined by the layer-dependent dynamical
self-energy $\Sigma _{D}(z,\omega )$. In bulk materials one distinguishes
Fermi liquid and Mott insulators by the low-frequency behavior of $\Sigma_{D}
$; in a Fermi liquid $\Sigma _{D}\rightarrow_{\omega \rightarrow 0}
(1-Z^{-1})\omega $ (leading to a quasiparticle with renormalized mass) while
in a Mott insulator $\Sigma _{D}\rightarrow_{\omega \rightarrow 0} \Delta
^{2}/\omega $ (leading to a gap in the spectrum). In the heterostructures we
study, we find that outside the high density region, correlations are weak ($%
Z \approx 1$), and that as one moves to the interior of thicker
heterostructures, correlations increase ($Z$ decreases). Mott insulating
solutions ($Z=0$) are never found; instead $\Sigma _{D}(z,\omega )\sim
\lbrack 1-Z^{-1}(z)]\omega $ with $0 < Z < 1$ for all layers $z$, although
in the interior of thick, large $U$ heterostructures $Z$ is only
nonvanishing because of leakage (quantum tuneling) of quasiparticles from
the edges, and goes exponentially towards zero.

The nonvanishing $Z$ indicates a Fermi liquid state with well defined
coherent quasiparticles (thus negligible low frequency scattering). In the
heterostructure context the quasiparticles form subbands, with quasiparticle
energies $E_{\alpha}(\vec k_{\parallel })$ and wave functions $\varphi
_{\alpha}(z;E_{\alpha}(\vec k_{\parallel}))$ which are the low energy
eigenfunctions and eigenvalues of 
\begin{equation}
\left[ Z^{-1}(z)E_{\alpha} \delta _{z,z^{\prime }}+\mu -H_{band}(\vec
k_{\parallel })-H_{Coul}\right] \varphi _{\alpha}(z^{\prime })=0 .
\label{qpeq}
\end{equation}%
(We note that the 2-site DMFT method used here is believed to give
reasonable results for $Z$ but of course neglects scattering effects. Near $%
E_{\alpha}=0$ scattering is unimportant but of course will increase at
higher energies.)

Numerical results for the coherent quasiparticles in a heterostructure with $%
n=3$ and $U=16t$ are shown in Fig.~\ref{fig:QP}. For these parameters we
find 8 quasiparticle bands with non-vanishing electron density. The
calculated dispersion relations are shown in panel (a) and are labelled $%
\alpha =1...8$ in order of decreasing electron density. We observe that the
band splittings depend on momentum because of the layer dependence of $Z$.
The corresponding quasiparticle weights $Z_{\alpha }$ and real-space wave
functions $\varphi _{\alpha }(z)$ at $\omega =0$ are shown in Fig.~\ref%
{fig:QP}(b) and (c,d), respectively. (These quantities vary somewhat over
the band also). $\ Z_{\alpha }$ is the smallest for the $\alpha =1$ subband
because its real-space wave function contains the largest weight at $z=\pm
0.5$, where the charge density is the largest and, therefore, the
correlation effect is the strongest [see Fig.~\ref{fig:QP}(c)]. $Z_{\alpha } 
$ generally increases with increasing $\alpha $, because as $\alpha $
increases the wave function amplitudes $|\varphi _{\alpha }(z)|$ decrease in
the high density regions (near $z=0$). The anomalies observed in $Z_{\alpha }
$ at $\alpha =4,6$ correspond to the increase of $|\varphi _{\alpha }(z=\pm
0.5)|$ due to the symmetry of the wave function [see Fig.~\ref{fig:QP}(d)].

\begin{figure}[tbp]
\epsfxsize=0.8\columnwidth \centerline{\epsffile{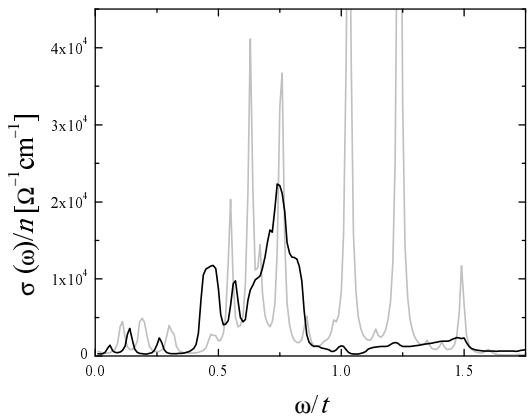}}
\caption{Heavy lines: low frequency (quasiparticle-region) optical
conductivity for $n=3$, $U=16t$ heterostructure. Peaks arise from transition
among subbands shown in Fig.~\protect\ref{fig:QP}(a). Light lines:
conductivity calculated from Hartree-Fock approximation to same Hamiltonian.}
\label{fig:optics}
\end{figure}

The coherent subbands may be studied by optical conductivity with electric
field directed along $[001]$. As an example, the heavy line in Fig.~\ref%
{fig:optics} shows the quasiparticle contribution to the conductivity
spectrum, calculated for a heterostructure with $n=3$ and $U=16t$ using the
standard Kubo formula with optical matrix element obtained by applying the
Peierls phase ansatz to $H_{band}$ (and $t=0.3$~eV). 
Three main features are evident at $\omega =0.75t, 0.55t$ and $0.45t$; 
each of these has contributions from two
interband transitions, ($1 \rightarrow 2$, $2 \rightarrow 3$), ($3
\rightarrow 4$, $4 \rightarrow 5$) and ($5 \rightarrow 8$, $6 \rightarrow 7$%
) respectively. The optical features are not sharp because the quasiparticle
band splitting depends on $\vec k_\parallel$. The weaker features at lower
energies arise from transitions involving high-lying, only slightly occupied
bands. The lighter lines in Fig.~\ref{fig:optics} show the optical
conductivity computed using the Hartree-Fock approximation. We see that the
spectra are qualitatively similar, but that the Hartree-Fock absorption
features occur at a larger energy because the $Z$-induced band narrowing is
absent and are delta functions because in the Hartree-Fock approximation the
subbands splittings are $\vec k_\parallel$-independent.

To summarize, we have presented the first dynamical mean field study of a
`correlated electron heterostructure', in which the behavior is controlled
by the spreading of the electronic charge out of the confinement region. Our
results show how the electronic behavior evolves from the weakly correlated
to the strongly correlated regions, and in particular confirms the existence
of an approximately three unit cell wide crossover region in which a system,
insulating in bulk, can sustain metallic behavior. We found that even in the
presence of very strong bulk correlations, the metallic edge behavior
displays a finite (roughly factor-of-two-to-three) 
mass renormalization. We showed
how the magnitude of the renormalization is affected by the spatial
structure of the quasiparticle wave function and determined how this
renormalization affects physical properties, in particular the optical
conductivity. Important future directions for research is to examine the
phase diagram, using beyond Hartree-Fock techniques, and to generalize the
results presented here to more complicated and realistic cases.

We acknowledge very fruitful discussions with M.~Potthoff and H.~Monien.
This research was supported by NSF DMR-00081075 (A.J.M.) and the JSPS (S.O.).

\end{document}